\pdfoutput=1
\RequirePackage{ifpdf}
\ifpdf 
\documentclass[pdftex]{sigma}
\else
\documentclass{sigma}
\fi

\newcommand{\bfX}{{\bf X}}
\newcommand{\bfN}{{\bf N}}

\newcommand{\maa}{{\mathcal A}}

\newcommand{\bfx}{{\bf x}}
\newcommand{\bfxp}{{{\bf x}^\prime}}
\newcommand{\N}{{\mathbf N}}

\newcommand{\R}{{\mathbf R}}
\newcommand{\Si}{{\mathbf S}}

\newcommand{\Z}{{\mathbf Z}}
\newcommand{\C}{{\mathbf C}}

\newcommand{\mcI}{{\mathcal I}}

\newcommand{\mcg}{{\mathcal G}}

\begin{document}


\renewcommand{\PaperNumber}{108}

\FirstPageHeading

\ShortArticleName{Opposite Antipodal Fundamental Solution of Laplace's Equation}

\ArticleName{Opposite Antipodal Fundamental Solution\\ of Laplace's Equation in Hyperspherical Geometry}

\Author{Howard S.~COHL $^{\dag\ddag}$}

\AuthorNameForHeading{H.S.~Cohl}

\Address{$^\dag$~Applied and Computational Mathematics Division, Information Technology Laboratory,\\
\hphantom{$^\dag$}~National Institute of Standards and Technology, Mission Viejo, California, 92694 USA}
\EmailD{\href{mailto:howard.cohl@nist.gov}{howard.cohl@nist.gov}}

\Address{$^\ddag$~Department of Mathematics, University of Auckland, 38 Princes Str., Auckland, New Zealand}

\ArticleDates{Received August 18, 2011, in final form November 22, 2011; Published online November 29, 2011}

\vspace{-1mm}

\Abstract{Due to the isotropy of $d$-dimensional hyperspherical space, one expects
there to exist a spherically symmetric opposite antipodal fundamental solution for its corresponding
Laplace--Beltrami operator. The $R$-radius hypersphere
${\mathbf S}_R^d$ with $R>0$, represents a~Riemannian manifold with
positive-constant sectional curvature. We obtain a spherically symmetric
opposite antipodal fundamental solution of Laplace's equation on this manifold in terms of its
geodesic radius. We give several matching expressions for this fundamental
solution including a definite integral over reciprocal powers of the trigonometric
sine, finite summation expressions over trigonometric functions, Gauss hypergeometric
functions, and in terms of the Ferrers function of the second with degree and order given
by $d/2-1$ and $1-d/2$ respectively, with argument $x\in(-1,1)$.}


\Classification{35A08; 35J05; 32Q10; 31C12; 33C05}
\vspace{-3mm}

\section{Introduction}
\label{Introduction}

\looseness=-1
We compute closed-form expressions of a spherically symmetric opposite antipodal Green's
function (opposite antipodal fundamental solution of Laplace's equation) for a $d$-dimensional Riemannian
manifold of positive-constant sectional curvature, namely the $R$-radius hypersphere
with $R>0$. This problem is intimately related to the solution of the Poisson equation on this manifold and the study of spherical harmonics
which play an important role in exploring collective motion of many-particle systems in
quantum mechanics, particularly nuclei, atoms and molecules. In these systems, the
hyperradius is constructed from appropriately mass-weighted quadratic forms from the
Cartesian coordinates of the particles. One then seeks either to identify discrete forms
of motion which occur primarily in the hyperradial coordinate, or alternatively to
construct complete basis sets on the hypersphere. This representation was introduced
in quantum mechanics by Zernike \& Brinkman~\cite{ZernBrink},
and later invoked to greater effect in nuclear and atomic physics, respectively,
by Delves~\cite{Delves} and Smith~\cite{Smith}. The relevance of this
representation to two-electron excited states of the helium atom was noted by
Cooper, Fano \& Prats~\cite{CooperFanoPrats}; Fock~\cite{Fock1,Fock2}
had previously shown that the hyperspherical representation was particularly
efficient in representing the helium wave function in the vicinity of small hyperradii.
There has been a rich literature of applications ever since. Examples include
Zhukov~\cite{Zhukov}
(nuclear structure), Fano~\cite{Fano}
and Lin~\cite{Lin} (atomic structure), and
Pack \& Parker~\cite{PackParker} (molecular collisions). A~recent
monograph by Berakdar~\cite{Berakdar} discusses hyperspherical harmonic methods
in the general context of highly-excited electronic systems.
Useful background material relevant for the mathematical aspects of this paper can be
found in~\cite{Lee,Thurston,Vilen}.
Some historical refe\-ren\-ces on this topic include~\cite{Higgs,Leemon,Schrodinger38,
Schrodinger40,VinMarPogSisStr}.

This paper is organized as follows.
In Section~\ref{Thehyperboloidmodelofhyperbolicgeometry},
we describe hyperspherical geometry and its corresponding metric,
global geodesic distance function, Laplace--Beltrami operator (Laplacian),
and hyperspherical global geodesic polar coordinate systems
which parametrize points on this manifold.
In Section~\ref{AGreensfunctioninthehyperboloidmodel}, for hyperspherical
geometry, we show how to
compute `radial' harmonics in a geodesic polar coordinate system
and derive several alternative expressions for a `radial' opposite antipodal fundamental solution of
the Laplace's equation on the $R$-radius hypersphere.

Throughout this paper we rely on the following definitions.
For $a_1,a_2,\ldots\in\C$, if $i,j\in\Z$ and $j<i$ then
$\sum\limits_{n=i}^{j}a_n=0$ and $\prod\limits_{n=i}^ja_n=1$.
The set of natural numbers is given by $\N:=\{1,2,3,\ldots\}$, the set
$\N_0:=\{0,1,2,\ldots\}=\N\cup\{0\}$, the set of integers is given by
$\Z:=\{0,\pm 1,\pm 2,\ldots\}$, the sets of real and complex numbers are
given by $\R$ and $\C$ respectively.

\vspace{-2mm}

\section{Hyperspherical geometry}
\label{Thehyperboloidmodelofhyperbolicgeometry}

\vspace{-1mm}

The Euclidean inner product for $\R^{d+1}$ is given by
$
(\bfx,{\mathbf y})=x_0y_0+x_1y_1+\cdots+x_dy_d$.
The variety $(\bfx,\bfx)=x_0^2+x_1^2+\cdots+x_d^2=R^2$, for $\bfx\in\R^{d+1}$
and $R>0$,
defines the $R$-radius hypersphere $\Si_R^{d}$. We denote the unit radius
hypersphere by $\Si^d:=\Si_1^d$.
Hyperspherical space in $d$-dimensions, denoted by $\Si_R^d$,
is a maximally symmetric, simply connected, $d$-dimensional Riemannian manifold
with positive-constant sectional curvature
(given by $1/R^2$, see for instance~\cite[p.~148]{Lee}),
whereas Euclidean space~$\R^d$ equipped with the Pythagorean norm, is a
Riemannian manifold with zero sectional curvature.

Points on the $d$-dimensional hypersphere $\Si_R^d$ can be parametrized using
{\it subgroup-type coordinate systems}, i.e., those which correspond to a maximal
subgroup chain ${\rm O}(d)\supset \cdots$ (see for instance~\cite{IPSWa,IPSWc}).
The isometry group of the space $\Si_R^d$ is the orthogonal group ${\rm O}(d)$.
Hyperspherical space $\Si_R^d$, can be identified with
the quotient space ${\rm O}(d)/{\rm O}(d-1)$. The isometry group ${\rm O}(d)$ acts transitively on~$\Si_R^d$.
There exist se\-pa\-rable coordinate systems on the hypersphere, analogous to
parabolic coordinates in Euclidean space, which can not be constructed
using maximal subgroup chains.
{\it Polyspherical coordinates}, are coordinates which correspond
to the maximal subgroup chain given by ${\rm O}(d)\supset \cdots$.
What we will refer to as {\it standard hyperspherical coordinates},
correspond to the subgroup chain given by
${\rm O}(d)\supset {\rm O}(d-1)\supset \cdots \supset {\rm O}(2)$.
(For a thorough discussion of polyspherical coordinates
see \cite[Section~IX.5]{Vilen}.)
Polyspherical coordinates on $\Si^{d}_R$ all share the property that they are described by
$(d+1)$-variables: $R\in[0,\infty)$ plus $d$-angles each being given by the
values $[0,2\pi)$, $[0,\pi]$, $[-\pi/2,\pi/2]$ or $[0,\pi/2]$
(see~\cite{IPSWa, IPSWb}).

In our context, a useful subset of polyspherical coordinate are {\it geodesic polar coordinates}
$(\theta,{\mathbf {\widehat x}})$
(see for instance~\cite{Oprea}).
These coordinates, which parametrize points on $\Si_R^d$, have origin at
$O=(R,0,\ldots,0)\in\R^{d+1}$ and are given by a `radial' parameter $\theta\in[0,\pi]$ which parametrizes
points along a geodesic curve emanating from $O$ in a direction
$
{\mathbf {\widehat x}}
\in\Si^{d-1}$.
Geodesic polar coordinate systems partition $\Si_R^d$ into a family of
$(d-1)$-dimensional hyperspheres, each with a~`radius' $\theta:=\theta_d\in(0,\pi),$ on
which all possible hyperspherical coordinate systems
for~$\Si^{d-1}$ may be used
(see for instance~\cite{Vilen}). One then must also
consider the limiting case for $\theta=0,\pi$ to fill out all of $\Si_R^d$.
{\it Standard hyperspherical coordinates}
(see~\cite{KalMilPog,Olevskii})
are an example of geodesic polar coordinates, and are given by{\samepage
\begin{gather}
\begin{array}{@{}l}
x_0 = R\cos\theta,\\
x_1=R\sin\theta\cos\theta_{d-1},\\
x_2=R\sin\theta\sin\theta_{d-1}\cos\theta_{d-2},\\
\cdots\cdots\cdots\cdots\cdots\cdots\cdots\cdots\cdots \\
x_{d-2} = R\sin\theta\sin\theta_{d-1}\cdots\cos\theta_{2},\\
x_{d-1} = R\sin\theta\sin\theta_{d-1}\cdots\sin\theta_{2}\cos\phi,\\
x_{d} = R\sin\theta\sin\theta_{d-1}\cdots\sin\theta_{2}\sin\phi,
\end{array}
\label{standardhyp}
\end{gather}
 $\theta_i\in[0,\pi]$ for $i\in\{2,\ldots,d\}$, $\theta =\theta_{d}$,
and $\phi\in[0,2\pi)$.}

In order to study an opposite antipodal fundamental solution of Laplace's equation on the hypersphere, we need to
describe how one computes the geodesic distance in this space. Geodesic distances on
$\Si_R^d$ are simply given by arc lengths, angles between two arbitrary vectors, from the
origin in the ambient Euclidean space (see for instance~\cite[p.~82]{Lee}).
Any parametrization of the hypersphere~$\Si_R^d$,
must have $(\bfx,\bfx)=x_0^2+\cdots+x_d^2=R^2$, with $R>0$. The distance between two
points $\bfx,\bfxp\in\Si_R^d$ on the hypersphere is given by
\begin{equation}
d(\bfx,\bfxp)=R\gamma
=R\cos^{-1}\left(\frac{(\bfx,\bfxp)}{(\bfx,\bfx)(\bfxp,\bfxp)} \right)
=R\cos^{-1}\left(\frac{1}{R^2}(\bfx,\bfxp)\right).
\label{cosgamma}
\end{equation}
This is evident from the fact that the geodesics on $\Si_R^d$ are great circles,
i.e., intersections of $\Si_R^d$ with planes through the origin of the ambient Euclidean space,
with constant speed parametrizations.

In any geodesic polar coordinate system, the
geodesic distance between two points on the submanifold
is given by
\begin{equation}
d(\bfx,\bfxp)=R\cos^{-1}\left(\frac{1}{R^2}(\bfx,\bfxp)\right)
=R\cos^{-1}
\bigl( \cos \theta\cos\theta^\prime + \sin\theta\sin\theta^\prime\cos\gamma\bigr),
\label{diststandard}
\end{equation}
where $\gamma$ is the unique separation angle
given in each
polyspherical coordinate system used to parametrize points on $\Si^{d-1}$.
For instance, the separation angle $\gamma$ in standard hyperspherical coordinates
is given through
\begin{equation}
 \cos \gamma=\cos(\phi-\phi^\prime)
\prod_{i=1}^{d-2}\sin\theta_i{\sin\theta_i}^\prime
+\sum_{i=1}^{d-2}\cos\theta_i{\cos\theta_i}^\prime
\prod_{j=1}^{i-1}\sin\theta_j{\sin\theta_j}^\prime.
\label{prodform}
\end{equation}
Corresponding separation angle formulae for any hyperspherical
coordinate system used to parametrize points on $\Si^{d-1}$ can
be computed using (\ref{cosgamma}) and the associated formulae
for the appropriate inner-products.

One can also compute the Riemannian (volume) measure $d{\rm vol}_g$
(see for instance \cite[Section~3.4]{Grigor}),
invariant under the isometry group ${\rm SO}(d)$, of the
Riemannian manifold~$\Si_R^d$. For instance, in standard hyperspherical
coordinates~(\ref{standardhyp}) on $\Si_R^{d}$ the volume measure is given by
\begin{gather}
{\rm d}{\rm vol}_g=
R^d\sin^{d-1}\theta\,{\rm d}\theta\,{\rm d}\omega:=
R^d\sin^{d-1}\theta\,{\rm d}\theta\,
\sin^{d-2}\theta_{d-1}\cdots\sin\theta_2\, {\rm d}\theta_{1}\cdots {\rm d}\theta_{d-1}.
\label{eucsphmeasureinv}
\end{gather}
The distance $r\in[0,\infty)$ along a geodesic, measured from the origin, is
given by $r=\theta R$. To show that the above volume measure (\ref{eucsphmeasureinv})
reduces to the Euclidean volume measure at small distances (see for instance~\cite{KalMilPog}), we examine the limit of zero curvature, the flat-space limit.
In order to do this, we take the limit $\theta\to 0^+$ and $R\to\infty$ of the
volume measure~(\ref{eucsphmeasureinv}) which produces
\[
{\rm d}{\rm vol}_g\sim R^{d-1}\sin^{d-1}\left(\frac{r}{R}\right){\rm d}r\, {\rm d}\omega\sim r^{d-1}{\rm d}r\,{\rm d}\omega,
\]
which is the Euclidean measure on $\R^d$, expressed in standard Euclidean
hyperspherical coordinates. This measure is invariant under the Euclidean motion group ${\rm E}(d)$.

It will be useful below to express the Dirac delta distribution on $\Si_R^d$. The Dirac delta
distribution on the Riemannian manifold $\Si_R^d$ with metric $g$ is defined for an
open set $U\subset\Si_R^d$ with $\bfx,\bfxp\in\Si_R^d$ such that
\begin{gather}
\int_U\delta_g(\bfx,\bfxp) {\rm d}{\rm vol}_g =
 \begin{cases}
 1 & \mathrm{if}\ \bfxp\in U, \\
 0 & \mathrm{if}\ \bfxp\notin U.
\end{cases}
\label{defndiracdeltafunction}
\end{gather}
For instance, using (\ref{eucsphmeasureinv}) and
(\ref{defndiracdeltafunction}), in standard hyperspherical coordinates on
$\Si_R^{d}$ (\ref{standardhyp}), we see that the Dirac delta distribution is given by
\[
\delta_g(\bfx,\bfxp)=\frac{\delta(\theta-\theta^\prime)}{R^d\sin^{d-1}\theta^\prime}
\frac{\delta(\theta_1-\theta_1^\prime)\cdots\delta(\theta_{d-1}-\theta_{d-1}^\prime)}
{
\sin\theta_2^\prime
\cdots
\sin^{d-2}\theta_{d-1}^\prime
}.
\]

\subsection{The Laplace and Poisson equations on the hypersphere}

Parametrizations of a submanifold embedded in Euclidean space
can be given in terms of coordinate systems whose coordinates are {\it curvilinear}. These are
coordinates based on some transformation that converts the standard Cartesian
coordinates in the ambient space to a coordinate system with the same number of
coordinates as the dimension of the submanifold in which the coordinate lines are curved.

The Laplace--Beltrami operator
(Laplacian) in curvilinear coordinates
$\boldsymbol{\xi}=\big(\xi^1,\ldots,\xi^d\big)$ on a~Riemannian manifold is given by
\begin{gather}
\Delta=\sum_{i,j=1}^d\frac{1}{\sqrt{|g|}}
\frac{\partial}{\partial \xi^i}
\biggl(\sqrt{|g|}g^{ij}
\frac{\partial}{\partial \xi^j}
 \biggr),
\label{laplacebeltrami}
\end{gather}
where $|g|=|\det(g_{ij})|,$ the metric is given by
\begin{gather}
{\rm d}s^2=\sum_{i,j=1}^{d}g_{ij}{\rm d}\xi^i{\rm d}\xi^j,
\label{metric}
\end{gather}
and
\[
\sum_{i=1}^{d}g_{ki}g^{ij}=\delta_k^j,
\]
where $\delta_i^j\in\{0,1\}$ is the Kronecker delta
\begin{gather}
\delta_i^j:=
 \begin{cases}
 1 & \mathrm{if}\ i=j, \\
 0 & \mathrm{if}\ i\ne j,
\end{cases}
\label{Kronecker}
\end{gather}
for $i,j\in\Z$.
The relationship between the metric tensor $G_{ij}=\mathrm{diag}(1,\ldots,1)$ in
the ambient space and $g_{ij}$ of (\ref{laplacebeltrami}) and (\ref{metric}) is given~by
\[
g_{ij}({\mathbf{\xi}})=\sum_{k,l=0}^dG_{kl}\frac{\partial x^k}{\partial \xi^i}
\frac{\partial x^l}{\partial \xi^j}.
\]

The Riemannian metric in a geodesic polar coordinate system
on the submanifold $\Si_R^d$ is given~by
\begin{gather}
{\rm d}s^2=R^2\big({\rm d}\theta^2+\sin^2\theta\, {\rm d}\gamma^2\big),
\label{stanhypmetric}
\end{gather}
where an appropriate expression for $\gamma$ in a curvilinear coordinate system is given.
If one combines~(\ref{standardhyp}),
(\ref{prodform}),
(\ref{laplacebeltrami})
and (\ref{stanhypmetric}), then in a geodesic polar coordinate system,
Laplace's equation
on~$\Si_R^d$ is given by
\begin{gather}
\Delta f=\frac{1}{R^2}\left[\frac{\partial^2f}{\partial\theta^2}
+(d-1)\cot\theta\frac{\partial f}{\partial \theta}
+\frac{1}{\sin^2\theta} \Delta_{\Si^{d-1}}f\right]=0,
\label{genhyplap}
\end{gather}
where $\Delta_{\Si^{d-1}}$ is the corresponding Laplace--Beltrami operator
on $\Si^{d-1}$.

\bigskip
Consider Poisson's equation on $\Si_R^d$, $-\Delta u=\rho$ on a compact Riemannian manifold $M$
with boundary $\partial M$.
The divergence theorem on this manifold is given by (cf.~\cite[p.~43]{Lee})
\begin{equation}
\int_M \operatorname{div} \bfX\, {\rm d}V=\int_{\partial M} \langle \bfX,\bfN\rangle\, \mathrm{d}{\tilde V},
\label{divergencethm}
\end{equation}
where ${\rm d}V$ is the Riemannian volume measure on $M$, $\bfN$ is the outward unit normal
to $\partial M$, and ${\rm d}{\tilde V}$ is the Riemannian volume measure of the induced
metric on $\partial M$. If one invokes the divergence theorem on $\Si_R^d$ with regard
to Poisson's equation on this manifold using $\bfX=\nabla u$, then since
$\partial \Si_R^d=\varnothing$, one ascertains
\[
\int_{\Si_R^d}\rho \,\mathrm{d}V=0.
\]
Hence on $\Si_R^d$ (and on all compact manifolds without boundary), there does not exist a~source density
distribution $\rho$, satisfying Poisson's equation, with non-vanishing integral.
In fact, a~fundamental solution of Laplace's equation on~$\Si_R^d$
(see Theorem \ref{thmh1d} below), which has been pointed
out in \cite[Section 5.4]{Chapling16}, is actually the solution
to the Poisson equation whose inhomogeneous source distribution is given
by a point source at the origin and another with opposite sign, on
the opposite pole of the hypersphere (both modeled by Dirac delta distributions).

We define the {\it opposite antipodal} fundamental solution of
Laplace's equation $\maa_R^d(\bfx,\bfx')$, as the solution to
the following distributional partial differential equation
\begin{equation}
\label{eq3}
-\Delta\maa_R^d(\bfx,\bfxp)=\delta_g(\bfx,\bfxp)-\delta_g(-\bfx,\bfxp),
\end{equation}
where $g$ is the Riemannian metric on $\Si_R^d$
(e.g., (\ref{stanhypmetric}))
and
$\delta_g$ is the Dirac delta distribution on the manifold $\Si_R^d$.
The total integral over the entire manifold of this source distribution
vanishes. Therefore the solution to the resulting partial differential
equation must exist.
The opposite antipodal fundamental solution is the most
natural fundamental solution of Laplace's equation on a $d$-dimensional
$R$-radius hypersphere $\Si_R^d$ because (1) it is spherically symmetric
and (2) its density distribution is composed wholly of the minimum
number, two, of isolated Dirac delta distributions. Note that many other
composed density distribution may be assembled by collecting a finite number
of Dirac delta distributions, or by assembling an infinite number of
Dirac delta distributions over the manifold.

\section{An opposite antipodal Green's function on the hypersphere}
\label{AGreensfunctioninthehyperboloidmodel}

\subsection{Harmonics in geodesic polar coordinates}\label{SepVarStaHyp}

The harmonics (solutions to Laplace's equation) in a geodesic polar coordinate system are given in terms of
a `radial' solution (`radial' harmonics) multiplied by the angular solution
(angular harmonics).

Using polyspherical coordinates on $\Si^{d-1},$ one can compute the
normalized hyperspherical harmonics in this space by solving the Laplace equation
using separation of variables. This results in a general procedure
which, for instance, is given explicitly in~\cite{IPSWa, IPSWb}.
These angular harmonics are given as general expressions involving trigonometric functions,
Gegenbauer polynomials and Jacobi polynomials.
The angular harmonics are eigenfunctions of the Laplace--Beltrami operator
on $\Si^{d-1}$ which satisfy the following eigenvalue problem
(see for instance \cite[equa\-tion~(12.4) and Corollary~2 to Theorem~10.5]{Takeuchi})
\begin{gather}
\Delta_{\Si^{d-1}}Y_l^K
({\mathbf {\widehat x}})
=-l(l+d-2)Y_l^K({\mathbf {\widehat x}}),
\label{eq4}
\end{gather}
where
${\mathbf {\widehat x}}\in\Si^{d-1}$,
$Y_l^K({\mathbf {\widehat x}})$
are normalized angular hyperspherical harmonics,
$l\in\N_0$ is the
angular momentum quantum number, and
$K$ stands for the set of $(d-2)$-quantum numbers identifying
degenerate harmonics for each $l$ and $d$.
The degeneracy
\[
(2l+d-2)\frac{(d-3+l)!}{l!(d-2)!}
\]
(see~\cite[equation~(9.2.11)]{Vilen}),
tells you how many linearly independent solutions exist for a particular $l$ value and dimension $d$.
The angular hyperspherical harmonics are normalized such that
\[
\int_{\Si^{d-1}} Y_l^K
({\mathbf {\widehat x}})
\overline{Y_{l^\prime}^{K^\prime}
({\mathbf {\widehat x}})
}
{\rm d}\omega=
\delta_{l}^{l^\prime}
\delta_{K}^{K^\prime},
\]
where ${\rm d}\omega$ is the Riemannian (volume) measure on $\Si^{d-1}$, which is invariant under
the isometry group ${\rm SO}(d)$
(cf.~(\ref{eucsphmeasureinv})), and for $x+{\rm i}y=z\in\C$, $\overline{z}=x-{\rm i}y$, represents complex conjugation.
The angular solutions (hyperspherical harmonics) are well-known
(see~\cite[Chapter~IX]{Vilen} and \cite[Chapter~11]{ErdelyiHTFII}).
The generalized Kronecker delta symbol~$\delta_K^{K^\prime}$
(cf.~(\ref{Kronecker}))
is defined such that it equals 1 if
all of the $(d-2)$-quantum numbers identifying degenerate harmonics for each~$l$ and~$d$ coincide,
and equals zero otherwise.

We now focus on the `radial' solutions (harmonics) on $\mathbf S_R^d$. These
satisfy the following ordinary differential equation
(cf.~(\ref{genhyplap}) and (\ref{eq4}))
\begin{gather}
\frac{{\rm d}^2u}{{\rm d}\theta^2}+(d-1)\cot\theta\frac{{\rm d}u}{{\rm d}\theta}-\frac{l(l+d-2)}{\sin^2\theta}u=0.
\label{sphericallysymmetricharmonicequation}
\end{gather}
Four solutions of this ordinary differential equation
$u_{1,\pm}^{d,l},u_{2,\pm}^{d,l}\colon (-1,1)\to\C$ are given by
\begin{gather*}
{u_{1,\pm}^{d,l}(\cos\theta):=\frac{1}{(\sin\theta)^{d/2-1}}{\sf P}_{d/2-1}^{\pm(d/2-1+l)}
(\pm\cos\theta)},
\end{gather*}
 and
\begin{gather}
u_{2,\pm}^{d,l}(\cos\theta):=\frac{1}{(\sin\theta)^{d/2-1}}{\sf Q}_{d/2-1}^{\pm(d/2-1+l)}
(\pm\cos\theta) ,
\label{u2pmdl}
\end{gather}
where
${\sf P}_\nu^\mu, {\sf Q}_\nu^\mu\colon (-1,1)\to\C$ are Ferrers functions of the first
and second kind.
The Ferrers functions of the first and second kind
(see~\cite[Chapter~14]{NIST:DLMF}) can be defined respectively in terms
of a sum over two Gauss hypergeometric functions,
for all $\nu,\mu\in\C$ such that $\nu+\mu\not\in-\N$,
\begin{gather*}
{\sf P}_\nu^\mu(x) :=
\frac{2^{\mu+1}}{\sqrt{\pi}}
\sin\left[\frac{\pi}{2}(\nu+\mu)\right]
\frac
{\Gamma\left(\frac{\nu+\mu+2}{2}\right)}
{\Gamma\left(\frac{\nu-\mu+1}{2}\right)}
x\big(1-x^2\big)^{-\mu/2}{}_2F_1
\left(\frac{1-\nu-\mu}{2},\frac{\nu-\mu+2}{2};\frac32;x^2\right)\\
\phantom{{\sf P}_\nu^\mu(x) := }{} +
\frac{2^{\mu}}{\sqrt{\pi}}
\cos\left[\frac{\pi}{2}(\nu+\mu)\right]
\frac
{\Gamma\left(\frac{\nu+\mu+1}{2}\right)}
{\Gamma\left(\frac{\nu-\mu+2}{2}\right)}
\big(1-x^2\big)^{-\mu/2}{}_2F_1
\left(\frac{-\nu-\mu}{2},\frac{\nu-\mu+1}{2};\frac12;x^2\right)
\end{gather*}
(cf.~\cite[equation~(14.3.11)]{NIST:DLMF}), and
\begin{gather}
{\sf Q}_\nu^\mu(x) :=
\sqrt{\pi}2^{\mu}\cos\left[\frac{\pi}{2}(\nu+\mu)\right]
\frac
{\Gamma\!\left(\frac{\nu+\mu+2}{2}\right)}
{\Gamma\!\left(\frac{\nu-\mu+1}{2}\right)}
x\big(1-x^2\big)^{-\mu/2}{}_2F_1
\left(\frac{1-\nu-\mu}{2},\frac{\nu-\mu+2}{2};\frac32;x^2\right)\!\nonumber\\
{} -\sqrt{\pi}2^{\mu-1}\sin\left[\frac{\pi}{2}(\nu+\mu)\right]
\frac
{\Gamma\left(\frac{\nu+\mu+1}{2}\right)}
{\Gamma\left(\frac{\nu-\mu+2}{2}\right)}
\big(1-x^2\big)^{-\mu/2}{}_2F_1
\left(\frac{-\nu-\mu}{2},\frac{\nu-\mu+1}{2};\frac12;x^2\right)
\label{ferrerssecondkinddefnhypergeom}
\end{gather}
(cf.~\cite[equation~(14.3.12)]{NIST:DLMF}).
The Gauss hypergeometric function
${}_2F_1\colon \C\times\C\times(\C\setminus-\N_0)\times \mathbf C\setminus[1,\infty)\to\mathbf C$,
can be defined in terms of the infinite series
\[
{}_{2}F_1(a,b;c;z):=
\sum_{n=0}^\infty \frac{(a)_n(b)_n}{(c)_n n!}z^n
\]
(see~\cite[equation~(15.2.1)]{NIST:DLMF}),
and elsewhere in $z$ by analytic continuation. On the unit circle \mbox{$|z|=1$}, the Gauss
hypergeometric series converges absolutely if $\operatorname{Re}(c-a-b)\in(0,\infty),$
converges conditionally if $z\ne 1$ and $\operatorname{Re}(c-a-b)\in(-1,0],$
and diverges if $\operatorname{Re}(c-a-b)\in(-\infty,-1]$. For $z\in\C$ and $n\in\N_0$,
the Pochhammer symbol $(z)_n$ (also referred to as the rising factorial) is defined as
(cf.~\cite[equation~(5.2.4)]{NIST:DLMF})
\[
(z)_n:=\prod_{i=1}^n(z+i-1).
\]
The Pochhammer symbol is expressible in terms of a quotient of gamma functions as \cite[equation~(5.2.5)]{NIST:DLMF}
\[
(z)_n=\frac{\Gamma(z+n)}{\Gamma(z)},
\]
for all $z\in\C\setminus-\N_0$.
The gamma function $\Gamma\colon \C\setminus-\N_0\to\C$
(see~\cite[Chapter~5]{NIST:DLMF}) is an important combinatoric
function and is ubiquitous in special function theory. It is naturally defined over
the right-half complex plane through Euler's integral
(see~\cite[equation~(5.2.1)]{NIST:DLMF})
\[
\Gamma(z):=\int_0^\infty t^{z-1} {\rm e}^{-t} {\rm d}t,
\]
 $\operatorname{Re} z>0$, and elsewhere by analytic continuation. The Euler reflection formula allows one to obtain values
of the gamma function in the left-half complex
plane \cite[equation~(5.5.3)]{NIST:DLMF},
namely
\[
\Gamma(z)\Gamma(1-z)=\frac{\pi}{\sin\pi z},
\]
$0<\operatorname{Re} z<1,$ for $\operatorname{Re}z=0$, $z\ne 0$, and then
for $z$ shifted by integers using
the following recurrence relation (see~\cite[equation~(5.5.1)]{NIST:DLMF})
\[
\Gamma(z+1)=z\Gamma(z).
\]
An important formula which the gamma function satisfies is the duplication
formula \cite[equation~(5.5.5)]{NIST:DLMF}
\begin{gather}
\Gamma(2z)=\frac{2^{2z-1}}{\sqrt{\pi}}\Gamma(z)\Gamma\left(z+\frac12\right),
\label{duplicationformulagamma}
\end{gather}
provided $2z\not\in-\N_0$.

Due to the fact that the space $\Si_R^d$ is homogeneous with respect to
its isometry group, the orthogonal group ${\rm O}(d)$, and therefore
an isotropic manifold, we expect that
there exist a fundamental solution on this space with spherically symmetric
dependence. We specifically expect these solutions to be given in terms of
the Ferrers function of the second kind with argument given by $\cos\theta$.
The Ferrers function of the second kind naturally fits our requirements
because it is
singular at $\theta=0$, whereas the Ferrers function of the first kind,
with the same argument, is regular at $\theta=0$. We require there to exist a
singularity at the origin of an opposite antipodal fundamental solution of Laplace's equation on~$\Si^d_R$, since it is a manifold and must behave locally like a Euclidean
fundamental solution of Laplace's equation which also has a singularity at the origin.

\subsection[Opposite antipodal fundamental solution of $-\Delta u=0$ on $\Si_R^d$]{Opposite antipodal fundamental solution of $\boldsymbol{-\Delta u=0}$ on $\boldsymbol{\Si_R^d}$}
\label{FunSolLapHd}

In computing an opposite antipodal fundamental solution of the Laplacian on $\Si_R^d$, we must solve~(\ref{eq3})
where $g$ is the Riemannian metric on $\Si_R^d$
(e.g., (\ref{stanhypmetric}))
and
$\delta_g$ is the Dirac delta distribution on the manifold $\Si_R^d$
(e.g., (\ref{defndiracdeltafunction})).
In general since we can add any harmonic function to an opposite antipodal fundamental solution of Laplace's
equation and still have an opposite antipodal fundamental solution, we will use this freedom to make our fundamental
solution as simple as possible. It is reasonable to expect that there exists a particular
spherically symmetric fundamental solution $\maa_R^d$ on the hypersphere
with pure `radial', $\theta:=d(\widehat{\bf x},\widehat{\bf x}')$ (e.g., (\ref{diststandard})),
and constant angular
dependence
due to the influence of the point-like nature of the Dirac delta distribution in (\ref{eq3}).
For a spherically symmetric solution to the Laplace equation, the corresponding
$\Delta_{\Si^{d-1}}$ term in (\ref{genhyplap})
vanishes since only the $l=0$ term survives in
(\ref{sphericallysymmetricharmonicequation}).
In other words we expect there to exist an opposite antipodal fundamental solution
of Laplace's equation on~$\Si_R^d$ such
that $\maa_R^d(\bfx,\bfxp)=f(\theta)$
(cf.~(\ref{diststandard})), where $R$ is a parameter of this fundamental solution.

We will prove that on the $R$-radius hypersphere $\Si_R^d$, an opposite
antipodal Green's function for
the Laplace operator (fundamental solution of Laplace's equation) can be given as follows.

\begin{theorem}\label{thmh1d}
Let $d\in\{2,3,\ldots\}$. Define $\mcI_d\colon (0,\pi)\to\R$ as
\[
\mcI_d(\theta):=\int_\theta^{\pi/2}\frac{{\rm d}x}{\sin^{d-1}x},
\]
${{\bf x}},{{{\bf x}^\prime}}\in\Si_R^d$, and
$\maa_R^d\colon \big(\Si_R^d\times\Si_R^d\big)\setminus\big\{(\bfx,\bfx)\colon \bfx\in\Si_R^d\big\}\to\R$ defined such that
\[
\maa_R^d({\bf x},{\bf x}^\prime):=
 \frac{\Gamma (d/2 )}{2\pi^{d/2}R^{d-2}}\mcI_d(\theta),
\]
where $\theta:=\cos^{-1}\left([{\widehat{\bf x}},{{\widehat{\bf x}^\prime}}]\right)$ is the geodesic distance between~${\widehat{\bf x}}$ and ${{\widehat{\bf x}^\prime}}$ on the unit radius hypersphere~$\Si^d$, with ${\widehat{\bf x}}=\bfx/R$, ${{\widehat{\bf x}^\prime}}=\bfxp/R$, then $\maa_R^d$ is an opposite antipodal fundamental solution for~$-\Delta$ where $\Delta$ is the Laplace--Beltrami operator on $\Si_R^d$. Moreover,
\begin{gather*}
\mcI_d(\theta) =
 \begin{cases}
\displaystyle \frac{(d-3)!!}{(d-2)!!}\biggl[\log\cot \frac{\theta}{2}
+\cos \theta\sum_{k=1}^{d/2-1}\frac{(2k-2)!!}{(2k-1)!!}\frac{1}{\sin^{2k}\theta}\biggr]
& \mathrm{if}\ d\ \mathrm{even}, \vspace{1mm}\\
\left\{ \begin{array}{l}
\displaystyle
\left(\frac{d-3}{2}\right)!
\sum_{k=1}^{(d-1)/2}
\frac{\cot^{2k-1}\theta}
{(2k-1)(k-1)!((d-2k-1)/2)!},
\nonumber\vspace{1mm}\\
\mathrm{or} \vspace{1mm}\\
\displaystyle
\frac{(d-3)!!}{(d-2)!!}\cos\theta
\sum_{k=1}^{(d-1)/2}
\frac{(2k-3)!!}{(2k-2)!!}
\frac{1}{\sin^{2k-1}\theta},
\end{array} \right\} & \mathrm{if}\ d\ \mathrm{odd},
\end{cases} \vspace{1mm}\\
\phantom{\mcI_d(\theta)}{}
 =
 \begin{cases}
 \displaystyle \cos\theta\
{}_2F_1\left(\frac12,\frac{d}{2};\frac{3}{2};\cos^2\theta\right),\vspace{1mm}\\
\displaystyle \frac{\cos\theta}{\sin^{d-2}\theta}\ {}_2F_1\left(1,\frac{3-d}{2};\frac32;\cos^2\theta\right),\vspace{1mm}\\
\displaystyle \frac{(d-2)!}{\displaystyle \Gamma\left(d/2\right)2^{d/2-1}}
\frac{1}{(\sin \theta)^{d/2-1}}{\sf Q}_{d/2-1}^{1-d/2}(\cos \theta).
\end{cases}
\end{gather*}
\end{theorem}

\begin{proof}
Since a spherically symmetric choice for an opposite antipodal fundamental
solution
satisfies Laplace's
equation everywhere except at $\theta\in\{0,\pi\}$, we may first set $g=f^\prime$
in (\ref{genhyplap}) and solve the first-order ordinary differential equation
\[
g^\prime+(d-1)\cos \theta\; g=0,
\]
which is integrable and clearly has the general solution
\begin{gather}
g(\theta)=\frac{{\rm d}f}{{\rm d}\theta}=c_0(\sin \theta)^{1-d},
\label{dfdr}
\end{gather}
where $c_0\in\R$ is a constant.
Now we integrate (\ref{dfdr}) to obtain an opposite antipodal fundamental
solution for the Laplacian on $\Si_R^d$
\begin{gather}
\maa_R^d(\bfx,\bfxp)=c_0\mcI_d(\theta)+c_1,
\label{Hdid01}
\end{gather}
where $\mcI_d\colon (0,\pi)\to\R$ is defined as
\begin{gather}
\mcI_d(\theta):=\int_\theta^{\pi/2}\frac{{\rm d}x}{\sin^{d-1}x},
\label{In}
\end{gather}
and $c_0,c_1\in\R$ are constants (independent of $\theta$) which depend on $d$ and $R$.
Notice that we can add any harmonic function to~(\ref{Hdid01}) and still have
an opposite antipodal fundamental solution of the Laplacian since an opposite antipodal fundamental solution of the
Laplacian must satisfy
\[
\int_{\Si_R^d} (-\Delta\varphi)(\bfxp) \maa_R^d(\bfx,\bfxp)\,{\rm d}{\rm vol}_g^\prime = \varphi(\bfx),
\]
for all $\varphi\in {\mathcal S}\big(\Si_R^d\big)$, where ${\mathcal S}$ is the space of test functions,
and $d{\rm vol}_g^\prime$ is the Riemannian (volume) measure on $\Si_R^d$ in the primed
coordinates.
Notice that our fundamental solution of Laplace's equation on the hypersphere~(\ref{Hdid01}),
has the property that it tends towards $+\infty$ as $\theta\to 0^{+}$ and
tends towards $-\infty$ as $\theta\to\pi^{-}$.
Therefore our fundamental solution attains all real values.
As an aside, by the definition therein
(see~\cite{Grigor83,Grigor87}),
$\Si_R^d$ is a parabolic manifold.
Since the hypersphere~$\Si_R^d$ is bi-hemispheric,
we expect that an opposite antipodal fundamental solution of Laplace's equation on the
hypersphere should vanish at $\theta=\pi/2$. It is therefore convenient
to set $c_1=0$ leaving us with
\begin{gather}
\maa_R^d(\bfx,\bfxp)=c_0\mcI_d(\theta).
\label{Hdid}
\end{gather}

In Euclidean space $\R^d$, a Green's function for Laplace's
equation (fundamental solution for the Laplacian)
is well-known and is given by the following expression
(see \cite[p.~94]{Fol3}, \cite[p.~17]{GT},
 \cite[p.~211]{BJS}, \cite[p.~6]{Doob}).
Let $d\in\N$. Define
\begin{gather}
\mcg^d({\bf x},{\bf x}^\prime)=
 \begin{cases}
\displaystyle\frac{\Gamma(d/2)}{2\pi^{d/2}(d-2)}\|{\bf x}-{\bf x}^\prime\|^{2-d}
& \mathrm{if}\ d=1\mathrm{\ or\ }d\ge 3,\vspace{1mm} \\
\displaystyle\frac{1}{2\pi}\log\|{\bf x}-{\bf x}^\prime\|^{-1}
& \mathrm{if}\ d=2,
\end{cases}
\label{thmg1n}
\end{gather}
then $\mcg^d$ is a fundamental solution
for $-\Delta$
in Euclidean space $\R^d$,
where $-\Delta$ is the positive Laplace operator on $\R^d$.
Note that most authors only present the above theorem for the case $d\ge 2$ but it is
easily-verified to also be valid for the case $d=1$ as well.

The hypersphere $\Si_R^d$, being a manifold, must behave locally like
Euclidean space $\R^d$. Therefore for small $\theta$ we have
${\rm e}^\theta\simeq 1+\theta$ and ${\rm e}^{-\theta}\simeq 1-\theta$ and in that limiting regime
\[
\mcI_d(\theta)\approx\int_\theta^1 \frac{{\rm d}x}{x^{d-1}}\simeq
 \begin{cases}
-\log \theta & \mathrm{if}\ d=2, \vspace{1mm} \\
{\displaystyle \frac{1}{\theta^{d-2}}} & \mathrm{if}\ d\ge 3, \end{cases}
\]
which has exactly the same singularity as a Euclidean fundamental solution.
Therefore the proportionality constant $c_0$ is obtained by matching locally
to a Euclidean fundamental solution{\samepage
\begin{gather}
\maa_R^d=c_0 \mcI_d\simeq {\mathcal G}^d,
\label{proportionality}
\end{gather}
in a small neighborhood of the singularity at $\bfx=\bfxp,$ as the curvature vanishes, i.e., $R\to\infty$.}

We have shown how to compute an opposite antipodal fundamental solution of the
Laplace--Beltrami operator on the hypersphere in terms of an improper integral~(\ref{In}). We now prove several equivalent
finite summation expressions for $\mcI_d(\theta)$. We
wish to compute the antiderivative $\mathfrak{I}_m\colon (0,\pi)\to\R$, which is defined to within a constant in~$x$ as
\[
\mathfrak{I}_m(x):=\int\frac{{\rm d}x}{\sin^mx},
\]
where $m\in\N$. This antiderivative satisfies the following recurrence relation
\begin{gather}
\mathfrak{I}_m(x)=-\frac{\cos x}{(m-1)\sin^{m-1}x}+\frac{(m-2)}{(m-1)}\mathfrak{I}_{m-2}(x),
\label{antiderivreccurence}
\end{gather}
which follows from the identity
\[
\frac{1}{\sin^{m}x}=\frac{1}{\sin^{m-2}x}+\frac{\cos x}{\sin^mx}\cos x,
\]
and integration by parts. The antiderivative $\mathfrak{I}_m(x)$ naturally breaks into
two separate classes, namely
\begin{gather}
\int\frac{{\rm d}x}{\sin^{2n+1}x}=-\frac{(2n-1)!!}{(2n)!!}
\left[\log\cot\frac{x}{2}+\cos x\sum_{k=1}^n\frac{(2k-2)!!}{(2k-1)!!}\frac{1}{\sin^{2k}x}\right]+C,
\label{antiderodd}
\end{gather}
and
\begin{gather}
\int\frac{{\rm d}x}{\sin^{2n}x}=
 \begin{cases}
\displaystyle
-\frac{(2n-2)!!}{(2n-1)!!}\cos x
\sum_{k=1}^n\frac{(2k-3)!!}{(2k-2)!!}\frac{1}{\sin^{2k-1}x}+C, \qquad\mbox{or}\vspace{1mm}\\
\displaystyle
-(n-1)!\sum_{k=1}^n\frac{\cot^{2k-1}x}{(2k-1)(k-1)!(n-k)!}+C,
\end{cases}
\label{antidereven}
\end{gather}
\noindent where $C$ is a constant.
The double factorial $(\cdot)!!\colon \{-1,0,1,\ldots\}\to\N$ is defined by
\[
n!!:=
 \begin{cases}
\displaystyle n\cdot(n-2)\cdots 2 & \mathrm{if}\ n\ \mathrm{even}\ge 2,
\nonumber \\
\displaystyle n\cdot(n-2)\cdots 1 & \mathrm{if}\ n\ \mathrm{odd}\ge 1,
 \\
\displaystyle 1 & \mathrm{if}\ n\in\{-1,0\}.
\end{cases}
\]
Note that $(2n)!!= 2^nn!$ for $n\in\N_0$.
The finite summation formulae for $\mathfrak{I}_m(x)$ all follow trivially by induction
using (\ref{antiderivreccurence}) and the binomial expansion
(cf.~\cite[equation~(1.2.2)]{NIST:DLMF})
\[
(1+\cos^2x)^n=n!\sum_{k=0}^n\frac{\cot^{2k}x}{k!(n-k)!}.
\]
The formulae (\ref{antiderodd}) and (\ref{antidereven}) are essentially
equivalent to \cite[equations~(2.515.1) and (2.515.2)]{Gradshteyn2007}, except
\cite[equation~(2.515.2)]{Gradshteyn2007} is in error
with the factor $28^k$ being replaced with $2^k$. This is also verified in
the original citing reference~\cite{Timofeev}. By applying
the limits of integration from the definition of~$\mcI_d(\theta)$ in~(\ref{In}) to
(\ref{antiderodd}) and (\ref{antidereven}) we obtain the following finite summation
expressions
\begin{gather}
\mcI_d(\theta)=
 \begin{cases}
\displaystyle \frac{(d-3)!!}{(d-2)!!}\left[\log\cot \frac{\theta}{2}
+\cos \theta\sum_{k=1}^{d/2-1}\frac{(2k-2)!!}{(2k-1)!!}\frac{1}{\sin^{2k}\theta}\right]
& \mathrm{if}\ d\ \mathrm{even}, \\
\left\{ \begin{array}{@{}l@{}}
\displaystyle
\left(\frac{d-3}{2}\right)!
\sum_{k=1}^{(d-1)/2}
\frac{\cot^{2k-1}\theta}
{(2k-1)(k-1)!((d-2k-1)/2)!},
\\
\mathrm{or} \\
\displaystyle
\frac{(d-3)!!}{(d-2)!!}\cos\theta
\sum_{k=1}^{(d-1)/2}
\frac{(2k-3)!!}{(2k-2)!!}
\frac{1}{\sin^{2k-1}\theta},
\end{array} \right\} & \mathrm{if}\ d\ \mathrm{odd}.
\end{cases}
\label{sumgradryzhikin}
\end{gather}

Moreover, the antiderivative (indefinite integral) can be given in terms of the Gauss
hypergeometric function as
\begin{gather}
\int\frac{{\rm d}\theta}{\sin^{d-1}\theta}=-\cos \theta
\; {}_2F_1\left(\frac12,\frac{d}{2};\frac32;\cos^2\theta\right)
+C,
\label{antiderivativecos2r}
\end{gather}
where $C\in\R$. This is verified as follows.
By using
\[
\frac{d}{dz}\; {}_2F_1(a,b;c;z)=\frac{ab}{c}\; {}_2F_1(a+1,b+1;c+1;z)
\]
(see~\cite[equation~(15.5.1)]{NIST:DLMF}),
and the chain rule, we can show that
\begin{gather*}
-\frac{{\rm d}}{{\rm d}\theta}
\cos\theta\;{}_2F_1\left(\frac12,\frac{d}{2};\frac32;\cos^2\theta\right)
\\
\qquad =
\sin\theta
\left[
{}_2F_1
\left(\frac12,\frac{d}{2};\frac32;\cos^2\theta\right)+\frac{d}{3}\cos^2\theta
\; {}_2F_1
\left(\frac32,\frac{d+2}{2};\frac52;\cos^2\theta\right)
\right].
\end{gather*}
The second hypergeometric function can be simplified using Gauss' relations for contiguous
hypergeometric functions, namely
\[
z\; {}_2F_1(a+1,b+1;c+1;z)=\frac{c}{a-b}\bigl[{}_2F_1(a,b+1;c;z)-{}_2F_1(a+1,b;c;z)\bigr]
\]
(see~\cite[p.~58]{Erdelyi}), and
\[
{}_2F_1(a,b+1;c;z)=\frac{b-a}{b}{}_2F_1(a,b;c;z)
+\frac{a}{b}\;{}_2F_1(a+1,b;c;z)
\]
(see~\cite[equation~(15.5.12)]{NIST:DLMF}).
By applying these formulae, the term with the hypergeometric function cancels leaving only a term
which is proportional to a binomial through
\[
{}_2F_1(a,b;b;z)=(1-z)^{-a}
\]
(see~\cite[equation~(15.4.6)]{NIST:DLMF}),
which reduces to $1/\sin^{d-1}\theta$. By applying
the limits of integration from the definition of $\mcI_d(\theta)$ in (\ref{In}) to~(\ref{antiderivativecos2r}), we obtain the following Gauss hypergeometric representation
\begin{gather}
\mcI_d(\theta)=
\cos\theta\;{}_2F_1\left(\frac12,\frac{d}{2};\frac32;\cos^2\theta\right).
\label{Idthetagausscos}
\end{gather}
Using (\ref{Idthetagausscos}), we can write another expression for $\mcI_d(\theta)$.
Applying Euler's transformation
\[
{}_2F_1(a,b;c;z)=(1-z)^{c-a-b}\; {}_2F_1\left(c-a,c-b;c;z\right)
\]
(see~\cite[equation~(2.2.7)]{AAR}),
to (\ref{Idthetagausscos}) produces
\[
\mcI_d(\theta)=
\frac{\cos\theta}{\sin^{d-2}\theta}\; {}_2F_1\left(1,\frac{3-d}{2};\frac32;\cos^2\theta\right).
\]

Our derivation for an opposite antipodal fundamental solution of Laplace's equation on the $R$-radius hypersphere
in terms of Ferrers functions of the second kind is as follows.
If we let $\nu+\mu=0$ in the definition of the Ferrers function of the second kind
${\sf Q}_\nu^\mu\colon (-1,1)\to\C$ (\ref{ferrerssecondkinddefnhypergeom}), we derive
\[
{\sf Q}_\nu^{-\nu}(x)=\frac{\sqrt{\pi}}{2^\nu}\frac{x\big(1-x^2\big)^{\nu/2}}
{\Gamma\left(\nu+\frac12\right)}\;
{}_2F_1\left(\frac12,\nu+1;\frac32;x^2\right),
\]
for all $\nu\in\C$. If we let $\nu=d/2-1$ and substitute $x=\cos\theta$, then we have
\begin{gather}
{\sf Q}_{d/2-1}^{1-d/2}(\cos\theta)=\frac{\sqrt{\pi}}{2^{d/2-1}}\frac{\cos\theta
\sin^{d/2-1}\theta}
{\Gamma\left(\frac{d-1}{2}\right)}\;
{}_2F_1\left(\frac12,\frac{d}{2};\frac32;\cos^2\theta\right).
\label{ferrerssecondnuminusnu}
\end{gather}
Using the duplication formula for gamma functions (\ref{duplicationformulagamma}),
then through (\ref{ferrerssecondnuminusnu}) we have
\begin{gather}
\mcI_d(\theta)=\frac{(d-2)!}{\Gamma(d/2)2^{d/2-1}}\frac{1}{\sin^{d/2-1}\theta}
{\sf Q}_{d/2-1}^{1-d/2}(\cos\theta).
\label{idferrerq2}
\end{gather}
We have therefore verified that the harmonics computed in
Section~\ref{SepVarStaHyp}, namely $u_{2,+}^{d,0}(\cos\theta)$ (\ref{u2pmdl}), give an alternate
form for an opposite antipodal fundamental solution of the Laplacian on the hypersphere.

The constant $c_0$ in an opposite antipodal fundamental solution for the Laplace operator
on the hypersphere~$\Si_R^d$~(\ref{Hdid}) is computed by locally matching up,
through (\ref{proportionality}), to the singularity
of an opposite antipodal fundamental solution for the Laplace operator in Euclidean space~(\ref{thmg1n}).
The coefficient $c_0$ depends on~$d$ and~$R$.
For $d\ge 3$ we take
the asymptotic expansion for $c_0\mcI_d(\theta)$ as $\theta\to 0^+$,
and match this to a fundamental solution of Laplace's equation for Euclidean space~(\ref{thmg1n}). This yields
\begin{gather}
\displaystyle c_0=\frac{\Gamma\left(d/2\right)}{2\pi^{d/2}}.
\label{c0gamma}
\end{gather}
For $d=2$ we take the asymptotic expansion for
\[
c_0\mcI_2(\theta)=-c_0\log\tan\frac{\theta}{2}\simeq
c_0\log\|\bfx-\bfxp\|^{-1},
\]
as $\theta\to 0^+$, and match this to
$\displaystyle \mcg^2(\bfx,\bfxp)=(2\pi)^{-1}\log\|\bfx-\bfxp\|^{-1},$
therefore $\displaystyle c_0=(2\pi)^{-1}$. This exactly
matches (\ref{c0gamma}) for $d=2$.
The $R$ dependence of $c_0$ originates from (\ref{In}), where $x$ and $\theta$
represents geodesic distances (cf.~(\ref{diststandard})).
The distance $r\in[0,\infty)$ along a geodesic, as measured from the origin
of $\Si_R^d$, is given by $r=\theta R$. To show that an opposite antipodal fundamental solution~(\ref{Hdid}) reduces to the Euclidean fundamental solution at small
distances (see for instance~\cite{KalMilPog}),
we examine the flat-space limit of zero curvature. In order to do this, we take the
limit $\theta\to 0^+$ and $R\to\infty$ of (\ref{In}) with the substitution
$x=r/R$ which produces a factor of~$R^{d-2}$. So an opposite antipodal fundamental solution
of Laplace's equation on the Riemannian manifold $\Si_R^d$ is given~by
\[
\maa_R^d({\bf x},{\bf x}^\prime):=
{\displaystyle \frac{\Gamma\left(d/2\right)}{2\pi^{d/2}R^{d-2}}\mcI_d\left(\theta\right)}.
\]
This completes the proof.
\end{proof}

Apart from the well-known historical results in two and three dimensions, the closed form
expressions for an opposite antipodal fundamental solution of Laplace's equation on the $R$-radius hypersphere
given by Theorem~\ref{thmh1d} in Section~\ref{FunSolLapHd} appear to be new. Furthermore, the
Ferrers function representations in Section~\ref{SepVarStaHyp} for the
radial harmonics on the $R$-radius hypersphere do not appear to be have previously appeared
in the literature.

\subsection[Some examples for $\mcI_d(\theta)$ and ${\sf Q}_{d/2-1}^{1-d/2}(\cos\theta)$]{Some examples for $\boldsymbol{\mcI_d(\theta)}$ and $\boldsymbol{{\sf Q}_{d/2-1}^{1-d/2}(\cos\theta)}$}
We would now like to express the integral $\mathcal I_d(\cos\theta)$ for $d\in\{2,\ldots,7\}$ in terms of
trigonometric and/or logarithmic functions.
The integral $\mcI_d$ can be computed using elementary
methods through its definition~(\ref{In}). In $d=2$ we have
\[
\mcI_2(\theta)=\int_\theta^{\pi/2} \frac{{\rm d}x}{\sin x}=
\frac{1}{2}\log \frac{\cos \theta+1}{\cos \theta-1}
=\log\cot\frac{\theta}{2},
\]
and in $d=3$ we have
\[
\mcI_3(\theta)=\int_\theta^{\pi/2} \frac{{\rm d}x}{\sin^2x}=\cot \theta.
\]
In $d\in\{4,5,6,7\}$ we have
\begin{gather*}
 \mcI_4(\theta) =
\frac12\log\cot\frac{\theta}{2}+
\frac{\cos \theta}{2\sin^2 \theta},\\
 \mcI_5(\theta) = \cot\theta+\frac13\cot^3\theta,\\
 \mcI_6(\theta)=
\frac38\log\cot\frac{\theta}{2}+\frac{3\cos\theta}{8\sin^2\theta}+\frac{\cos\theta}{4\sin^2\theta}
,\qquad\mathrm{and}\\
 \mcI_7(\theta) =
\cot\theta+\frac23\cot^3\theta+\frac15\cot^5\theta.
\end{gather*}
Equivalently, some relevant Ferrers
functions of the second kind ${\sf Q}_{d/2-1}^{1-d/2}(\cos\theta)$ for $d\in\{2,\ldots,7\}$ are
(cf.~(\ref{sumgradryzhikin}) and (\ref{idferrerq2}))
\begin{gather*}
{\sf Q}_0(\cos \theta) = \log\cot\frac{\theta}{2},\\
\frac{1}{(\sin\theta)^{1/2}}{\sf Q}_{1/2}^{-1/2}(\cos \theta) = \sqrt{\frac{\pi}{2}}\cot \theta,\\
\frac{1}{\sin\theta}{\sf Q}_1^{-1}(\cos\theta) =
\frac12\log\cot\frac{\theta}{2}+\frac{\cos\theta}{2\sin^2 \theta},\\
\frac{1}{(\sin\theta)^{3/2}}{\sf Q}_{3/2}^{-3/2}(\cos\theta) = \frac12\sqrt{\frac{\pi}{2}}
\left(\cot \theta+\frac13\cot^3\theta\right), \\
\frac{1}{(\sin\theta)^2}{\sf Q}_2^{-2}(\cos\theta) = \frac18\log\cot\frac{\theta}{2}+
\frac{\cos\theta}{8\sin^2\theta}+\frac{\cos\theta}{12\sin^4\theta},\qquad \mbox{and}
\\
\frac{1}{(\sin\theta)^{5/2}}{\sf Q}_{5/2}^{-5/2}(\cos\theta) = \frac18\sqrt{\frac{\pi}{2}}
\left(
\cot\theta+\frac23\cot^3\theta+\frac15\cot^5\theta
\right).
\end{gather*}

\subsection*{Acknowledgments}

Much thanks to Ernie Kalnins, Willard Miller~Jr., George Pogosyan, and Charles Clark for valuable discussions. Much thanks as well to Richard Chapling for his comments in~\cite{Chapling16}, in re\-fe\-rence to an original version of the present paper and its implications. I would like to express my sincere gratitude to the anonymous referees and an editor at SIGMA whose helpful comments improved this paper. This work was partly conducted while H.S.~Cohl was a National Research Council Research Postdoctoral Associate in the Information Technology Laboratory at the National Institute of Standards and Technology, Gaithersburg, Maryland, USA.

\pdfbookmark[1]{References}{ref}
\LastPageEnding


\begin{thebibliography}{99}
\footnotesize\itemsep=0pt

\bibitem{AAR}
 Andrews G.E., Askey R., Roy R.,
Special functions,
{\em Encyclopedia of Mathematics and its Applications}, Vol.~71,
 Cambridge University Press, Cambridge, 1999.

\bibitem{Berakdar}
Berakdar J.,
Concepts of highly excited electronic systems,
Wiley-VCH, New York, 2003.

\bibitem{BJS}
Bers L., John F., Schechter M. (Editors),
Partial differential equations,
Interscience Publishers, New York, 1964.

\bibitem{Chapling16}
Chapling R., A hypergeometric integral with applications to the fundamental
 solution of {L}aplace's equation on hyperspheres, \href{https://doi.org/10.3842/SIGMA.2016.079}{\textit{SIGMA}} \textbf{12} (2016),
 079, 20~pages, \href{https://arxiv.org/abs/1508.06689}{arXiv:1508.06689}.

\bibitem{CooperFanoPrats}
 Cooper J.W., Fano U., Prats F.,
Classification of two-electron excitation levels of helium,
\href{https://doi.org/10.1103/PhysRevLett.10.518}{{\em Phys. Rev. Lett.}} {\bf 10} (1963), 518--521.

\bibitem{Delves}
 Delves L.M.,
Tertiary and general-order collisions.~II,
\href{https://doi.org/10.1016/0029-5582(60)90174-7}{{\em Nuclear Phys.}} {\bf 20} (1960), 275--308.

\bibitem{Doob}
 Doob J.L.,
Classical potential theory and its probabilistic counterpart,
{\em Grundlehren der Mathematischen Wissenschaften}, Vol.~262,
Springer-Verlag, New York, 1984.

\bibitem{Erdelyi}
Erd{\'e}lyi A., Magnus W., Oberhettinger F., Tricomi F.G.,
Higher transcendental functions, Vol.~I,
Robert E. Krieger Publishing Co. Inc., Melbourne, Fla., 1981.

\bibitem{ErdelyiHTFII}
Erd{\'e}lyi A., Magnus W., Oberhettinger F., Tricomi F.G.,
Higher transcendental functions, Vol.~II,
Robert E. Krieger Publishing Co. Inc., Melbourne, Fla., 1981.

\bibitem{Fano}
Fano U.,
Correlations of two excited electrons,
\href{https://doi.org/10.1088/0034-4885/46/2/001}{{\em Rep. Progr. Phys.}} {\bf 46} (1983), 97--165.

\bibitem{Fock1}
Fock V.,
On the Schr\"odinger equation of the helium atom.~I,
{\it Norske Vid. Selsk. Forhdl.} {\bf 31} (1958), no.~22, 7~pages.

\bibitem{Fock2}
Fock V.,
On the Schr\"odinger equation of the helium atom.~II,
{\it Norske Vid. Selsk. Forhdl.} {\bf 31} (1958), no.~23, 8~pages.

\bibitem{Fol3}
 Folland G.B.,
Introduction to partial differential equations,
{\it Mathematical Notes}, Princeton University Press, Princeton, N.J., 1976.

\bibitem{GT}
Gilbarg D., Trudinger N.S.,
Elliptic partial differential equations of second order, 2nd ed.,
{\it Grundlehren der mathematischen Wissenschaften}, Vol.~224,
 Springer-Verlag, Berlin, 1983.

\bibitem{Gradshteyn2007}
Gradshteyn I.S., Ryzhik I.M.,
Table of integrals, series, and products, 7th ed., Elsevier/Academic Press, Amsterdam, 2007.

\bibitem{Grigor83}
Grigor'yan A.A.,
Existence of the Green's function on a manifold,
\href{https://doi.org/10.1070/RM1983v038n01ABEH003400}{{\em Russ. Math. Surv.}} {\bf 38} (1983), no.~1, 190--191.

\bibitem{Grigor87}
Grigor'yan A.A., On the existence of positive fundamental solutions of the
 {L}aplace equation on {R}iemannian manifolds, \href{https://doi.org/10.1070/SM1987v056n02ABEH003040}{\textit{Math. USSR Sb.}} \textbf{56} (1987), 349--358.

\bibitem{Grigor}
Grigor'yan A.A.,
Heat kernel and analysis on manifolds,
{\it AMS/IP Studies in Advanced Mathematics}, Vol.~47,
Amer. Math. Soc., Providence, RI, International Press, Boston, MA, 2009.

\bibitem{Higgs}
 Higgs P.W.,
Dynamical symmetries in a spherical geometry.~I,
\href{https://doi.org/10.1088/0305-4470/12/3/006}{{\em J.~Phys.~A: Math. Gen.}} {\bf 12} (1979), 309--323.

\bibitem{IPSWa}
 Izmest'ev A.A., Pogosyan G.S., Sissakian A.N., Winternitz P.,
Contractions of Lie algebras and separation of variables. The $n$-dimensional sphere,
\href{https://doi.org/10.1063/1.532820}{{\em J.~Math. Phys.}} {\bf 40} (1999), 1549--1573.

\bibitem{IPSWb}
 Izmest'ev A.A., Pogosyan G.S., Sissakian A.N., Winternitz P.,
Contractions of Lie algebras and the separation of variables: interbase expansions,
\href{https://doi.org/10.1088/0305-4470/34/3/314}{{\em J.~Phys.~A: Math. Gen.}} {\bf 34} (2001), 521--554.

\bibitem{IPSWc}
 Izmest'ev A.A., Pogosyan G.S., Sissakian A.N., Winternitz P.,
Contraction and interbases expansions on $n$-sphere,
in \href{https://doi.org/10.1142/9789812777850_0047}{Quantum Theory and Symmetries} (Krak\'ow, 2001),
 World Sci. Publ., River Edge, NJ, 2002, 389--395.

\bibitem{KalMilPog}
Kalnins E.G., Miller W. Jr., Pogosyan G.S.,
The Coulomb-oscillator relation on $n$-dimensional spheres and hyperboloids,
\href{https://doi.org/10.1134/1.1490116}{{\em Phys. Atomic Nuclei}} {\bf 65} (2002), 1086--1094.

\bibitem{Lee}
 Lee J.M.,
Riemannian manifolds,
{\em Graduate Texts in Mathematics}, Vol.~176,
 Springer-Verlag, New York, 1997.

\bibitem{Leemon}
 Leemon H.I.,
Dynamical symmetries in a spherical geometry.~II,
\href{https://doi.org/10.1088/0305-4470/12/4/009}{{\em J.~Phys.~A: Math. Gen.}} {\bf 12} (1979), 489--501.

\bibitem{Lin}
 Lin C.D.,
Hyperspherical coordinate approach to atomic and other Coulombic three-body systems,
\href{https://doi.org/10.1016/0370-1573(94)00094-J}{{\em Phys. Rep.}} {\bf 257} (1995), 1--83.

\bibitem{Olevskii}
 Olevski{\u\i} M.N.,
Triorthogonal systems in spaces of constant curvature in which the equation $\Delta_2u+\lambda u=0$ allows a complete separation of variables,
{\em Mat. Sbornik N.S.} {\bf 27} (1950), 379--426.


\bibitem{NIST:DLMF}
Olver F.W.J., Olde Daalhuis A.B., Lozier D.W., Schneider B.I., Boisvert R.F., Clark C.W., Miller B.R., Saunders B.V. (Editors),
{NIST} digital library of mathematical functions, {R}elease 1.0.21 of 2018-12-15, available at \url{http://dlmf.nist.gov}.

\bibitem{Oprea}
Oprea J.,
Differential geometry and its applications, 2nd ed.,
{\it Classroom Resource Materials Series}, Mathematical Association of America, Washington, DC, 2007.

\bibitem{PackParker}
 Pack R.T., Parker G.A.,
Quantum reactive scattering in 3 dimensions using hyperspherical (APH) coordinates. Theory,
\href{https://doi.org/10.1063/1.452944}{{\em J.~Chem. Phys.}} {\bf 87} (1987), 3888--3921.

\bibitem{Schrodinger38}
Schr{\"o}dinger E.,
Eigenschwingungen des sph\"arischen Raumes,
 {\em Comment. Pontificia Acad. Sci.} {\bf 2} (1938), 321--364.

\bibitem{Schrodinger40}
Schr{\"o}dinger E.,
A method of determining quantum-mechanical eigenvalues and eigenfunctions,
{\em Proc. Roy. Irish Acad. Sect.~A} {\bf 46} (1940), 9--16.

\bibitem{Smith}
 Smith F.T.,
Generalized angular momentum in many-body collisions,
\href{https://doi.org/10.1103/PhysRev.120.1058}{{\em Phys. Rev.}} {\bf 120} (1960), 1058--1069.

\bibitem{Takeuchi}
Takeuchi M.,
Modern spherical functions,
{\em Translations of Mathematical Monographs}, Vol.~135,
Amer. Math. Soc., Providence, RI, 1994.

\bibitem{Thurston}
Thurston W.P.,
Three-dimensional geometry and topology, Vol.~1,
 {\em Princeton Mathematical Series}, Vol.~35,
Princeton University Press, Princeton, NJ, 1997.

\bibitem{Timofeev}
Timofeev A.F.,
Integration of functions,
 OGIZ, Moscow~-- Leningrad, 1948 (in Russian).

\bibitem{Vilen}
Vilenkin N.Ja.,
Special functions and the theory of group representations,
 {\it Translations of Mathematical Monographs}, Vol.~22, Amer. Math. Soc., Providence, R.I., 1968.

\bibitem{VinMarPogSisStr}
Vinitski\u{\i} S.I., Mardoyan L.G., Pogosyan G.S., Sissakian A.N., Strizh T.A.,
Hydrogen atom in curved space. Expansion in free solutions on a three-dimensional sphere,
{\em Phys. Atomic Nuclei} {\bf 56} (1993), 321--327.

\bibitem{ZernBrink}
Zernike F., Brinkman H.C.,
Hypersph\"arische Funktionen und die in sph\"arische Bereichen orthogonalen Polynome,
{\em Proc. Akad. Wet. Amsterdam} {\bf 38} (1935), 161--170.

\bibitem{Zhukov}
 Zhukov M.V., Danilin B.V., Fedorov D.V., Bang J.M., Thompson I.J., Vaagen J.S.,
Bound state properties of Borromean halo nuclei: $^6$He and $^{11}$Li,
\href{https://doi.org/10.1016/0370-1573(93)90141-Y}{{\em Phys. Rep.}} {\bf 231} (1993), 151--199.

\end{thebibliography}
\end{document}